\documentclass[manuscript]{aastex}

\shorttitle{Observations of a Mass-Loading Eruption}
\shortauthors{Seaton et al.}

\begin{document}

\title{SWAP-SECCHI Observations of a Mass-Loading Type Solar Eruption}

\author{Daniel B. Seaton, Marilena Mierla \altaffilmark{1,2}, David Berghmans, Andrei N. Zhukov \altaffilmark{3, 4}, \and Laurent Dolla \altaffilmark{3}}
\affil{SIDC-Royal Observatory of Belgium, Avenue Circulaire 3, 1180 Brussels, Belgium}
\email{dseaton@oma.be}

\altaffiltext{1}{Institute of Geodynamics of the Romanian Academy, Bucharest, Romania}
\altaffiltext{2}{Research Center for Atomic Physics and Astrophysics, Faculty of Physics, University of Bucharest, Romania}
\altaffiltext{3}{Solar-Terrestrial Center of Excellence-SIDC, Brussels, Belgium}
\altaffiltext{4}{Skobeltsyn Institute of Nuclear Physics, Moscow State University, 119992 Moscow, Russia}

\begin{abstract}
We present a three-dimensional reconstruction of an eruption that occurred on 3~April~2010 using observations from \emph{SWAP} onboard \emph{PROBA2} and \emph{SECCHI} onboard \emph{STEREO}. The event unfolded in two parts: an initial flow of cooler material confined to a height low in the corona, followed by a flux rope eruption higher in the corona. We conclude that mass off-loading from the first part triggered a rise, and, subsequently, catastrophic loss of equilibrium of the flux rope.\end{abstract}


\section{Introduction}

Though Coronal Mass Ejections (CMEs) have been known to the scientific community for decades, the mechanisms responsible for their initiation remain the subject of much discussion. Numerous authors have proposed models that explain the physical processes that might trigger the destabilization of structures in the low corona and cause eruptions \citep[see the review of][]{Forbes2006}. 

In particular, a number of authors have studied the conditions under which a catastrophic loss-of-equilibrium may lead to an eruption \citep{PriestForbes2002}. \citet{ForbesIsenberg1991} proposed a two-dimensional analytic model describing how a coronal filament can erupt when its magnetic energy reaches a critical value. This loss-of-equilibrium model has been expanded and improved by a series of authors \citep{ForbesPriest1995, LinForbes2000, Reeves2006} who have shown how magnetic reconnection in the wake of such an eruption can accelerate a CME and produce a solar flare.  More recently, these two-dimensional, analytical models have been extended to three dimensions using numerical models by authors such as \citet{Roussev2003}, \citet{Kliem2004}, and \citet{Schrijver2008}.

In fact, there is essentially universal agreement that magnetic reconnection is the primary mechanism by which the stored magnetic energy necessary to accelerate a CME and generate a solar flare is released.  In most models an initial perturbation leads to the catastrophic loss-of-equilibrium of a flux rope and, therefore, an eruption. However, what mechanisms are responsible for that perturbation remains an open question.  One of the reasons for this lack of consensus is that there are a number of different ways in which a CME can be triggered. Flux cancellation and magnetohydrodynamic instabilities are common proposals, but some authors have explored additional mechanisms, including the role of mass loading and unloading.

\citet{WolfsonDlamini1997} and \citet{WolfsonSaran1998} found that excess mass loading could contribute to the stored energy necessary to lift an eruption against the Sun's gravity.  \citet{Low1996} and several others have suggested that a reduction of mass in a prominence can cause the prominence to rise buoyantly and thus lead to an eruption. \citet{Klimchuk2001}, however, rightly argues that given the number of CMEs that occur in the absence of a massive prominence, such a mechanism can only account for some subset of all eruptions.

One confounding factor in determining the cause of a CME is that there are relatively few examples where observers can clearly identify what mechanism actually initiated the event. However, in this Letter we present one such observation of an eruption: a hybrid of the mass-loading and loss-of-equilibrium types described above. The event, which occurred on 3~April~2010, triggered the CME that eventually caused the failure of the Galaxy 15 satellite.  In our analysis, the event begins when a mass off-load in an underlying filament triggers the slow rise of a coronal flux rope.  When the flux rope reaches a critical height, a catastrophic loss-of-equilibrium occurs and the flux rope erupts and produces a CME.

\section{Observations}\label{Observations}

Data from the \emph{Solar TErrestrial RElations Observatory (STEREO)} make it possible to use triangulation to derive the three-dimensional location of various solar phenomena: loops \citep[e.g.][]{Feng2007, Aschwanden2008, Rodriguez2009}; polar coronal jets \citep{Patsourakos2008}; solar filaments and prominences \citep{Gissot2008, Bemporad2009, Gosain2009, Liewer2009}; and CMEs \citep[see the review of][]{Mierla2010}.  In order to perform three-dimensional reconstructions of the features that make up this event, we must use a pair of images taken from different perspectives. In this case, we used pairs of images taken from two different perspectives of the \emph{STEREO} spacecraft. We used two different passbands from the \emph{Extreme UltraViolet Imager (EUVI)} on \emph{Sun Earth Connection Coronal and Heliospheric Investigation (SECCHI)}: 195~\AA\ (Fe~\textsc{xii} at $\log\,T \approx 6.17$) and 304~\AA\ (He~\textsc{ii} at $\log\,T \approx 4.9$). In our dataset, these pairs of \emph{EUVI}~195~\AA\ images from the two spacecraft were acquired with a cadence of 5~minutes, while the 304~\AA\ images were acquired with a cadence of 10~minutes. \citep[For a complete discussion of \emph{SECCHI} see][]{Howard2008}.

We also used observations from the \emph{Sun Watcher with Active Pixels and Image Processing (SWAP)} \citep{Berghmans2006, DeGroof2008, Seaton2010}, a wide-field extreme-ultraviolet (EUV) solar imager onboard the \emph{PROBA2} spacecraft in Earth orbit.  The \emph{SWAP} observations consist of a series of images at 174~\AA\ (Fe~\textsc{ix/x} at $\log\,T \approx 6.0$) taken with cadence of roughly 100 seconds.  \emph{SWAP} images are $1024 \times 1024$~pixels with a linear pixel size of approximately $3.17 \arcsec$, meaning the instrument has a total field of view of approximately $54 \times 54\;\mathrm{arcmin}^{2}$. (\emph{SWAP's} field of view is similar in size to \emph{EUVI's}, in contrast to the higher resolution, but smaller, field of view of \emph{AIA} on \emph{SDO}.)  In these images, we focused on a region of interest approximately $950\arcsec \times 950\arcsec$ in angular size, centered on NOAA~AR~11059.

Though \emph{EUVI} can acquire images in the 171~\AA\ passband, it observed at very low cadence in this channel on 3~April~2010 and did not acquire any images that included the eruption itself. Thus we were unable to use matched sets of images in the same passband from all three spacecraft.

\begin{figure*}

\begin{center}
  \includegraphics[scale=.8]{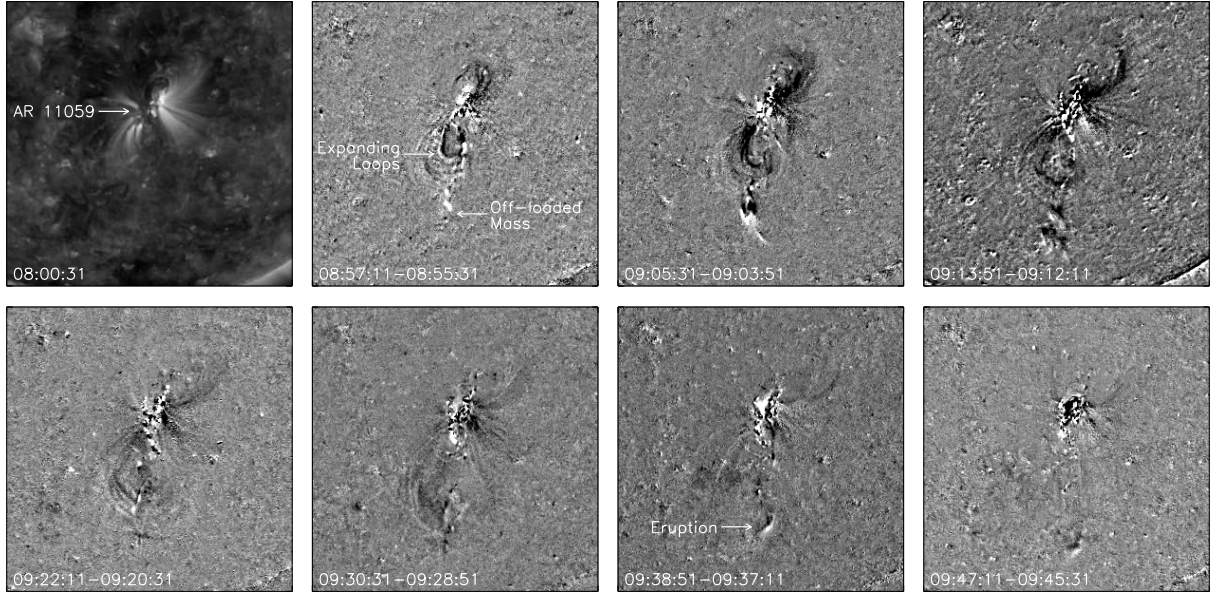}
  \caption{A series of running difference images from the \emph{SWAP} observations revealing the main features of the event.  Here we show every fourth image, with the preceding image (acquired approximately 100~s earlier) subtracted.  The first panel (a normal frame) provides an overview of the region of interest before the eruption began. In the second panel we indicate the large overlying loops, which expand during the early phases of the event to allow the flux rope's rise, as well as the initial mass flow that triggered the eruption.  In the next to last panel we indicate the final flow that occurred during the eruption, which appears to be composed mainly of the remnants of the erupted flux rope.\label{fig1}}

\end{center}
\end{figure*}

Figure~\ref{fig1} shows an overview of the event as seen by \emph{SWAP}.  The first panel shows an image of our region of interest obtained at about UT~08:00, before the eruption began. At this point, the region was dominated by a dark, S-shaped structure at the center of a series of large overlying loops. Overlaying the \emph{SWAP} images of this dark structure on magnetograms from \emph{GONG} revealed that it ran roughly along the magnetic neutral line of an essentially bipolar active region.  The north end of the dark feature was anchored in a region of positive polarity, while the south end was anchored in a region of negative polarity.

The subsequent panels of Figure~\ref{fig1} show running-difference images of the evolution of the region as the eruption unfolded, highlighting changes that occurred between pairs of sequential images. In the figure, each subsequent panel is separated by 500~s from the one preceding it, while the individual frames used to compose each panel are separated by 100~s.  These images reveal the main features of the eruption: in the second panel we see the initial flow of mass that triggered the event, which we refer to as the mass off-load.  In this and the next several frames we can also see the expansion of the overlying loops that followed the initial mass flow. By about UT~09:30 the eruption was well underway and, in the corresponding panels, we see the core of the eruption, which traveled southward through the \emph{SWAP} field of view as the event unfolded. 

At the time of the event, the \emph{STEREO} spacecraft were separated by $138.63^{\circ}$, with \emph{STEREO-A} leading the earth by $67.43^{\circ}$ and \emph{STEREO-B} trailing by $71.19^{\circ}$, meaning that objects near disk-center in \emph{SWAP} appear on the east limb in images from \emph{STEREO-A} and the west limb in images from \emph{STEREO-B}.  From these \emph{SECCHI} images, which have approximately twice the spatial resolution of \emph{SWAP} images, we identified the subregions corresponding to our \emph{SWAP} region-of-interest for analysis.  These subregions are shown in the top row of images in Figure~\ref{fig2}, which we will discuss in detail in the next section.

\section{Reconstruction and Analysis}\label{Analysis}

In order to understand the relationships between the individual features that made up this event, it is useful to know where they were in three-dimensional space.  For example, \emph{SWAP} images, because they show the eruption from almost directly above, tell us very little about how high above the solar surface any individual element was.  Thus we used \emph{SECCHI} images to make a series of three-dimensional reconstructions of the structures involved in the eruption.

To make these reconstructions we used a triangulation technique described in \citet{Inhester2006}. In this approach, we treat the positions of the two spacecraft used in the reconstruction as two viewpoints or observers.  The two positions of these spacecraft and the point in the solar corona to be triangulated define a plane called the epipolar plane. Since every epipolar plane is seen from a point lying inside that plane by both spacecraft, it is reduced to a single line---called the epipolar line---in the respective image projections. Any object identified to be situated on a certain epipolar line in one image must lie on the corresponding epipolar line in the other image.

Finding a correspondence between pixels in each image therefore requires only matching structures that appear along the correct epipolar line in each image.  Once we identify the feature in each image, we can make the three-dimensional reconstruction by tracking the lines of sight for each feature back into three-dimensional space.  Since these lines of sight must lie in the same epipolar plane, their intersection defines a unique location in space. Thus we can establish the three-dimensional position of any feature that appears in both images.

We used the \emph{SolarSoft} routine \textbf{scc\_measure.pro} to reconstruct features seen in either 195~\AA\ and 304~\AA\ images from \emph{SECCHI}. Once we had determined the three-dimensional locations of points within these structures, we could also determine how they should appear when seen in projection in images from all three instruments---that is, both of the \emph{SECCHI} instruments as well as \emph{SWAP}.  

\begin{figure*}

\begin{center}
  \includegraphics[scale=1]{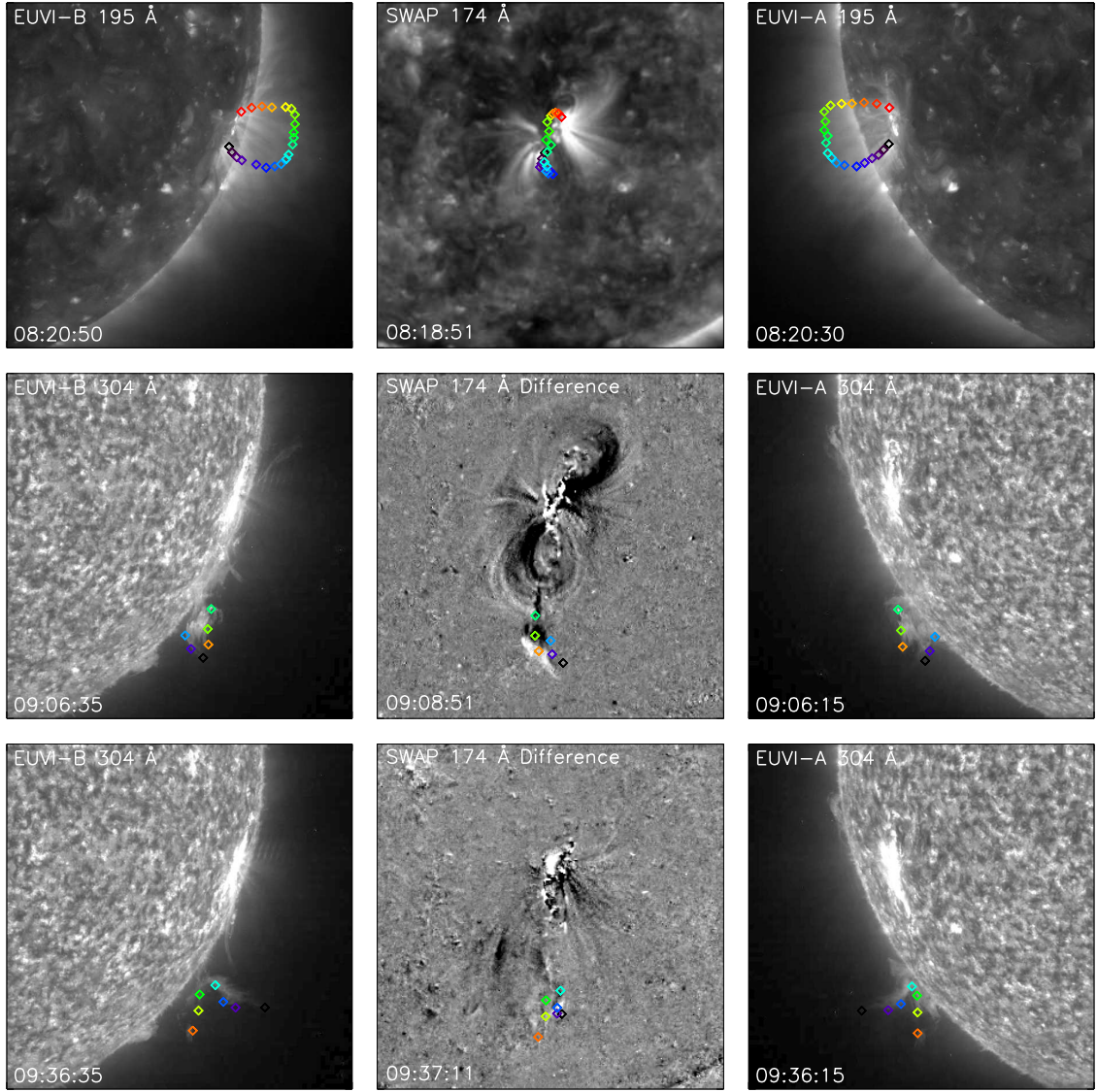}
  \caption{Reconstructions of three different structures seen during the eruption projected onto views from (left to right) \emph{EUVI-B}, \emph{SWAP}, and \emph{EUVI-A}.  The first row shows a reconstruction of the large, loop-like structure whose rise we track in the upper panel of figure~\ref{fig3}. The second row shows the initial mass flow that triggered the eruption as it moves southward. This flow is tracked in group ``B'' in  figure~\ref{fig3}. The final row shows our reconstruction of the final flow of the eruption, group ``C'' in figure~\ref{fig3}. (Note that we use color to help identify individual reconstructed points in projections from each point of view.) \label{fig2}}
\end{center}
\end{figure*}

We used the 195~\AA\ images to study the structure and height of a large, faint, loop-like feature---presumably the top of the erupting flux rope---that appeared to be associated with the active region.  This feature (seen at the beginning of the event) is shown in the top three panels of Figure~\ref{fig2}.  We have plotted the projected positions of the reconstructed points as they appear from the perspective of each spacecraft's viewpoint on top of the respective images.

\begin{figure*}
\begin{center}
  \includegraphics[scale=1]{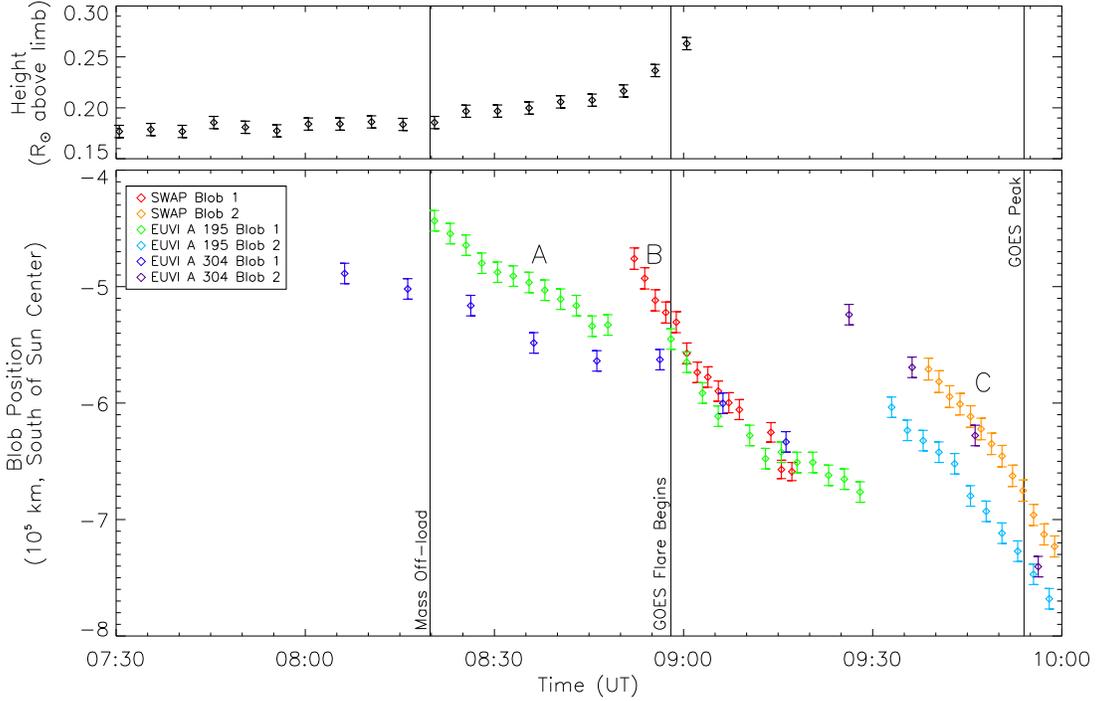}
  \caption{Height-time and position-time plots showing the movement of several structures observed during the eruption. The upper panel shows the rise of the large flux rope using points obtained from three-dimensional reconstructions of that feature.  The lower panel shows three flows of material as they travel southward during the eruption measured in projection in images from either \emph{SWAP} or \emph{EUVI-A}.  Group ``A'' represents the earliest flow from the active region (which is not visible in \emph{SWAP}), while group ``B'' shows a second flow that was visible in both \emph{EUVI} passbands and \emph{SWAP}. Group ``C'' indicates a later flow, which first appeared further south and higher in the corona than the earlier flows. This material appears to be mostly the remains of the erupted flux rope.\label{fig3}}
\end{center}
\end{figure*}

Using reconstructions for each of the \emph{SECCHI} images in which this feature is visible, we were able to track its maximum height throughout the eruption. The upper panel in Figure~\ref{fig3} shows these measured heights as a function of time. The figure shows that the loop structure remains relatively stable until the appearance of the mass-offload in the \emph{SECCHI} 195~\AA\ passband around UT~08:20, when it slowly begins to rise.  This rise accelerates gradually until UT~09:00, when the entire structure rapidly erupts and the top of the structure leaves the \emph{SECCHI} field of view. Since the front of the outgoing CME appears in \emph{COR1} coronagraph images from \emph{SECCHI} at UT~09:10, it seems likely that this feature forms part of the CME associated with the event.

We made two additional reconstructions using 304~\AA\ \emph{SECCHI} images as well.  The second row of Figure~\ref{fig2} shows the first of these, of the blob of material that comprised the initial mass-offload as it travels south from AR~11059.  Projections of the reconstructed points onto \emph{SWAP} images confirm that this structure is the same as the initial blob seen in running difference images of the event (i.e., the structure visible in the third panel of Figure~\ref{fig1}).  This blob travels southward for about one hour before it is overtaken by a larger blob of material that is directly associated with the eruption.  We show a reconstruction of this second blob in the third set of panels in Figure~\ref{fig2}.

We tracked the southward progress of both of these blobs as seen in projection from \emph{SECCHI-A} and \emph{SWAP}, using a combination of passbands and display techniques, and plotted their southward progress in the bottom panel of Figure~\ref{fig3}.  Using \emph{SWAP} images, we tracked two blobs of material that were clearly visible in running-difference movies.  Using \emph{SECCHI} images, we tracked two blobs in both the 304~\AA\ passband and 195~\AA| passband (using running-difference images in the latter case).  Because \emph{PROBA2} and \emph{STEREO-A} do not orbit the sun in the same plane, the viewing angle for the two spacecraft differs by several degrees.  To account for this difference, we applied a correction to the measured, projected positions of features seen in \emph{SWAP} images to match them to those seen in the \emph{SECCHI} images.

After plotting the locations of the individual blobs we could detect using the different imagers and bandpasses, we identified three groups of plotted points that appeared to be related.  The first, labeled ``A'' in Figure~\ref{fig3}, is an initial flow of material seen from \emph{SECCHI} that originates close to the center of the AR~11059.  The second, labeled ``B'', essentially a second pulse of flow that appears behind the initial one, is visible to both imagers.  The third, labeled ``C'', first appears further south, after the initial flow from the active region seems to have subsided. The blob associated with the points we see in group C appears to be composed mainly of the remnants of the erupting flux rope.

Comparing the timing of the appearance of these three flows to the rise of the large loop-like structure in the top panel of Figure~\ref{fig3} revealed that flow~A and the beginning of the slow rise of the loop-like structure occurred at approximately the same time. Flow~B occurred just as the slowly rising loop began to accelerate, and is associated with the structures we reconstructed in the second panel of Figure~\ref{fig2}.  Apparently both of these flows are part of the mass off-loading process that precipitated the eruption.

Figure~\ref{fig3} also shows that flow~A does not appear to be co-located in the two \emph{SECCHI} channels we studied, while flow~B appears in all three observations in essentially the same location.  Apparently the first pulse consisted of mostly cool, dense material, seen predominantly in the 304~\AA\ channel---it is only visible in running-difference images in the 195~\AA\ channel---while the subsequent flows were hotter, and therefore more clearly visible in all three channels we studied. In particular, we note that flow~B is bright in the view from \emph{SWAP}, which is a clear indication that some heating has occurred. This heating appears to be entirely confined to the filamentary material, since we see no additional brightening in \emph{SWAP} images for several minutes.

Once the rising loop began to accelerate, it rapidly left the \emph{EUVI} field-of-view. Minutes later, \emph{GOES} detected the first X-ray emission from the active region, signifying the onset of the flare (this is marked in Figure~\ref{fig3} by a vertical line).  The fact that the X-ray flare began at approximately the same time as the erupting loop structure underwent rapid acceleration is a clear indicator that the magnetic reconnection that released the energy necessary to accelerate the CME and generate flare emission \citep{Reeves2006} only began later in the event. Thus we conclude that reconnection is only responsible for accelerating the erupting structures, rather than triggering the event. The initial rise apparently occurred in a quasi-equilibrium state and was due to the increased buoyancy of the erupting flux rope because of the mass off-loading that occurred as a result of flows~A and B.  Only once this equilibrium breaks down completely does the eruption begin in earnest.

Flow~C, which is associated with the features we reconstructed in the bottom panel of Figure~\ref{fig3}, first appears at a location further south than either of the first two flows and travels southward until it reaches the limb of the sun as seen by \emph{SWAP}, at which point it effectively disappears from view.  From our reconstructions, we know that this feature is generally higher in the corona than flow~B, and that it does not appear to follow the same trajectory as flow~B either.  This latter flow appears to be associated with the remains of the erupted flux rope, rather than the initial off-loaded mass flow.

\section{Discussion} 

While models that might explain the dynamics and initiation of solar eruptions are numerous \citep{Klimchuk2001, Forbes2006}, observations that clearly show the mechanism of CME initiation are relatively rare. The event we present in this paper appears to be one of these rare observations: we see clear evidence that the eruption trigger is of the mass-loading type, and the subsequent evolution of the event appears to match the loss-of-equilibrium model.  (Another example of such an observation, featuring a different initiation mechanism, is described in \citealp{SterMoore2005}.)

One question we cannot answer is what caused the initial flow of mass from the active region. Images from both \emph{EUVI} and \emph{SWAP} reveal evidence of instability in the underlying filament in the hours before the eruption, particularly to the south of the active region. However, the active region itself remained essentially unchanged during this time and we see little evidence of heating or impulsive flows---as has been observed in other eruptions like the one discussed in \citet{Bone2009}---that might signal that magnetic reconnection was taking place before the large mass flow occurred.

The fact that the mass flow appears bright in \emph{SWAP} suggests that some localized heating did occur in the filament during the end of the off-load phase of the eruption, but the mechanism for this heating is impossible to identify. Additionally, neither EUV light curves from \emph{LYRA} onboard \emph{PROBA2} nor X-ray light curves from \emph{GOES} show any increase in brightness in the first 30~minutes after the first mass flow occurs. If reconnection had played a significant role in lifting the flux rope during the first phase of the event, we would expect some evidence of large-scale heating to appear in these light curves. Thus, while we cannot conclude that reconnection played no part in triggering or heating the mass flow, we can say with some certainty that reconnection was not a driver of the early phase of the eruption.

That this event appears to be triggered and driven by a combination of processes described by different CME initiation models \citep[e.g.][]{LinForbes2000, Low1996} is not surprising. Eruptions can begin in active regions with very different initial magnetic configurations, so it is likely that individual events will have different trigger mechanisms. This fact suggests that there is no silver-bullet model that will describe the initiation of all eruptions; a single event cannot resolve a question as long-standing and complex as what causes eruptions.

\acknowledgments

We are grateful to S.~Koutchmy for helpful discussions. Support for this paper came from PRODEX grant No. C90345 managed by the European Space Agency in collaboration with the Belgian Federal Science Policy Office (BELSPO) in support of the PROBA2/SWAP mission, and from the European Commission's Seventh Framework Programme (FP7/2007-2013) under the grant agreement No. 218816 (SOTERIA project, www.soteria-space.eu). \emph{SWAP} and \emph{LYRA} are projects of the Centre Spatial de Liege and the Royal Observatory of Belgium funded by the Belgian Federal Science Policy Office (BELSPO). We are thankful for the anonymous referee's thoughtful response, which helped us improve the paper.

%% To help institutions obtain information on the effectiveness of their
%% telescopes, the AAS Journals has created a group of keywords for telescope
%% facilities. A common set of keywords will make these types of searches
%% significantly easier and more accurate. In addition, they will also be
%% useful in linking papers together which utilize the same telescopes
%% within the framework of the National Virtual Observatory.
%% See the AASTeX Web site at http://www.journals.uchicago.edu/AAS/AASTeX
%% for information on obtaining the facility keywords.

%% After the acknowledgments section, use the following syntax and the
%% \facility{} macro to list the keywords of facilities used in the research
%% for the paper.  Each keyword will be checked against the master list during
%% copy editing.  Individual instruments or configurations can be provided 
%% in parentheses, after the keyword, but they will not be verified.

{\it Facilities:} \facility{PROBA2 (SWAP)}, \facility{STEREO (EUVI)}, \facility{GONG}

\begin{singlespace}

\end{singlespace}

\clearpage

\clearpage

\clearpage

%% The following command ends your manuscript. LaTeX will ignore any text
%% that appears after it.

\end{document}